\documentclass[conference,10pt]{IEEEtran}
\IEEEoverridecommandlockouts

\usepackage{cite}
\usepackage{amsmath,amssymb,amsfonts}
\usepackage{algorithmic}
\usepackage{graphicx}
\usepackage{textcomp}
\usepackage{xcolor}
\def\BibTeX{{\rm B\kern-.05em{\sc i\kern-.025em b}\kern-.08em
    T\kern-.1667em\lower.7ex\hbox{E}\kern-.125emX}}
\usepackage{comment}

\usepackage{booktabs}
\usepackage{tabularx}
\usepackage[table]{xcolor}
\usepackage{makecell}
\usepackage{adjustbox}
\usepackage[normalem]{ulem} 

\newcommand{\bestblock}[1]{\uline{#1}} 

\newcommand{\uarup}{\makebox[1.2em][l]{$\uparrow$}}
\newcommand{\uardown}{\makebox[1.2em][l]{$\downarrow$}}

\newcommand{\uarblank}{\makebox[1.2em][l]{}}
\usepackage{booktabs,tabularx,makecell,caption}
\begin{document}

\title{Prompt Amplification and Zero-Shot Late Fusion in Audio-Language Models for Speech Emotion Recognition
}

\author{\IEEEauthorblockN{Saurabh Kataria}
\IEEEauthorblockA{\textit{School of Nursing} \\
\textit{Emory University}\\
Atlanta, USA \\
saurabhkataria10@gmail.com}
\and
\IEEEauthorblockN{Xiao Hu}
\IEEEauthorblockA{\textit{School of Nursing} \\
\textit{Emory University}\\
Atlanta, USA \\
xiao.hu@emory.edu}
}


\maketitle

\begin{abstract}
Audio-Language Models (ALMs) are making strides in understanding speech and non-speech audio.
However, domain-specialist Foundation Models (FMs) remain the best for closed-ended speech processing tasks such as Speech Emotion Recognition (SER).
Using ALMs for Zero-shot SER is a popular choice, but their potential to work with specialists to achieve state-of-the-art (SOTA) performance remains unexplored.
We propose ZS-Fuse, a late-fusion method that combines zero-shot emotion estimates from a dual-encoder ALM with specialist FMs.
To handle ambiguity in emotions and sensitivity to prompt choice, 1) we use a simple prompt ensemble and 2) suggest a novel technique called prompt amplification, which repeats audio and text queries to discover stronger zero-shot capabilities.
We demonstrate the efficacy of our technique by evaluating ZS-Fuse with three dual-encoder ALMs and two FMs, and report improvements over SOTA baselines, such as WavLM-Large, on three speech emotion recognition datasets.
\end{abstract}

\begin{IEEEkeywords}
Speech Emotion Recognition, Audio-Language Models, Zero-Shot, Late Fusion, Prompt Engineering
\end{IEEEkeywords}

\section{Introduction}
Speech Emotion Recognition (SER) is a crucial paralinguistic task in human understanding~\cite{schuller2013interspeech}.
It facilitates the development of emotion-aware spoken dialogue systems~\cite{ma2020survey}, empathetic assistants~\cite{ma2023emotion}, customer satisfaction~\cite{parra2022classification}, and mental state monitoring~\cite{jordan2025speech}.
Recent works such as EmoBox~\cite{ma2024emobox} and VoxEmo~\cite{zhang2026voxemo} demonstrate the growing potential of Foundation Models (FM) and Audio-Language Models (ALM) for SER.

\begin{figure*}[t]
    \centering
    \includegraphics[width=0.96\linewidth]{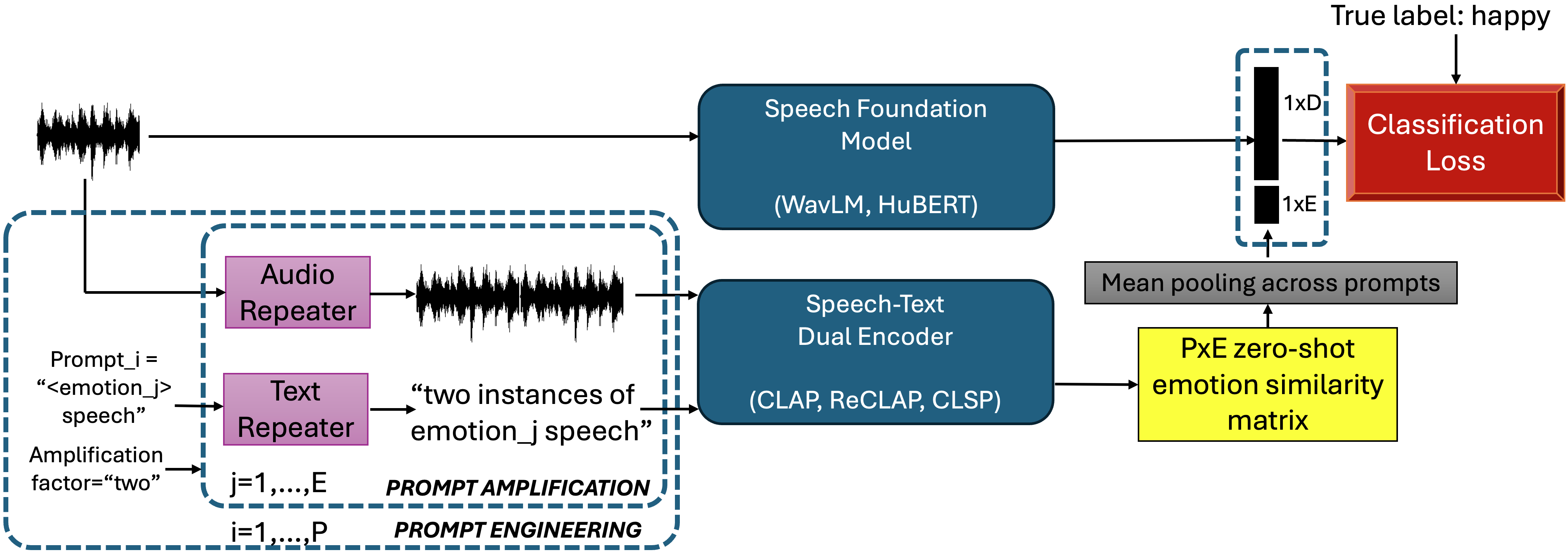}
    \caption{Proposed Zero-shot Fusion (ZS-Fuse) method. The speech foundation fine-tuning branch is complemented by zero-shot estimates from an audio-text dual encoder. The similarity score ($P\times E$) of pre-determined prompt templates ($P$) is computed for all emotion classes ($E$). Additionally, \textit{prompt amplification} trick can be used to increase the audio length by repeating it and prompt strength by replacing ``\texttt{<emotion\_j>}'' with ``\texttt{N instances of <emotion\_j>}'', for further potential gains.}
    \label{fig:zsfuse}
\end{figure*}

While conventional FMs achieve state-of-the-art (SOTA) performance~\cite{zhang2026voxemo,zhao2025steering}, ALMs provide semantic grounding, reasoning, and zero-shot prediction capabilities.
However, they suffer from over-confidence, which is addressed via confidence calibration, retrieval-based grounding, and uncertainty estimation~\cite{huang2025survey}.
Our proposal, ZS-Fuse (Fig.~\ref{fig:zsfuse}), safely navigates this issue by using zero-shot scores as a semantic prior and late fusing them with the conventional supervised FM fine-tuning pipeline.
We choose the Dual-Encoder (DE) ALM family over generative ones, such as Audio Flamingo 3~\cite{goel2025audio} and Qwen 3~\cite{xu2025qwen3}, because it aligns closely with our closed-set (fixed number of emotion classes) setup.
Contrastive Language-Audio Pretraining (CLAP)~\cite{elizalde2023clap} is a prominent instance of DE ALM, trained on caption-audio pairs and is suitable for open-vocabulary retrieval and comparing arbitrary text-audio pairs.

A related work, ABHINAYA~\cite{dutta2025abhinaya}, uses FMs and generative ALMs, but does decision-level fusion.
\cite{zhao2025steering} created a generative ALM using multiple FMs, but overall does not consistently surpass SOTA.
There have been prior efforts to develop paralinguistic FMs (emotion2vec~\cite{ma2024emotion2vec}) and DE ALMs (ParaCLAP~\cite{jing2024paraclap}), but they do not examine their combination.
Hence, in our proposal ZS-Fuse, 1) we focus specifically on surpassing SOTA and propose a simple late fusion method to combine predictions of dual-encoder Audio-Language Models with Foundation Model representations, 2) we propose a novel technique, \textit{prompt amplification}, which repeats audio and text queries to discover stronger zero-shot capabilities of ALMs, 3) we report improvements over the strong WavLM-Large baseline and discuss potential pitfalls in zero-shot fusion.

\begin{table*}[!t]
\centering
\caption{
UAR (mean$\pm$std) across three datasets.
Bold=global best for each dataset. Underline=best within each backbone block and per dataset.
The first row shows the random baseline.
For each backbone block, arrows indicate a change relative to the corresponding \texttt{+ none} baseline.
The last column indicates the number of results that improved ($\uparrow$) or degraded ($\downarrow$) when using an Audio-Language Model.
}
\label{tab:baseline}
\setlength{\tabcolsep}{2.2pt}
\renewcommand{\arraystretch}{1.08}
\begin{adjustbox}{max width=0.97\textwidth}
\small
\begin{tabularx}{\textwidth}{@{}>{\raggedright\arraybackslash}p{3.25cm} >{\centering\arraybackslash}X >{\centering\arraybackslash}X >{\centering\arraybackslash}X >{\centering\arraybackslash}p{1.75cm}@{}}
\toprule
\textbf{Setting (FM + ALM)} & \textbf{RAVDESS UAR} & \textbf{MSP-Podcast UAR} & \textbf{IEMOCAP UAR} & \textbf{Imp. Count} \\
\midrule
\rowcolor{black!6}
\makecell[l]{\textbf{Random}} &
0.2628$\pm$0.0450\uarblank &
0.1241$\pm$0.0106\uarblank &
0.2503$\pm$0.0017\uarblank &
-- \\
\midrule
\rowcolor{black!4}\multicolumn{5}{@{}l@{}}{\textbf{FM 1.1: WavLM-Base+}}\\
\makecell[l]{WavLM-Base+ + none}   & 0.6953$\pm$0.0994\uarblank & \bestblock{0.2955$\pm$0.0037}\uarblank & 0.6736$\pm$0.0060\uarblank & baseline \\
\makecell[l]{WavLM-Base+ + clap}   & 0.6719$\pm$0.0442\uardown & 0.2945$\pm$0.0003\uardown & 0.7023$\pm$0.0183\uarup & 1$\uparrow$, 2$\downarrow$ \\
\makecell[l]{WavLM-Base+ + reclap} & \bestblock{0.7969$\pm$0.0442}\uarup & 0.2862$\pm$0.0026\uardown & 0.6833$\pm$0.0190\uarup & 2$\uparrow$, 1$\downarrow$ \\
\makecell[l]{WavLM-Base+ + clsp}   & 0.7500$\pm$0.0442\uarup & 0.2946$\pm$0.0060\uardown & \bestblock{0.7045$\pm$0.0093}\uarup & 2$\uparrow$, 1$\downarrow$ \\
\midrule
\rowcolor{black!4}\multicolumn{5}{@{}l@{}}{\textbf{FM 1.2: WavLM-Large}}\\
\makecell[l]{WavLM-Large + none}   & 0.8438$\pm$0.0552\uarblank & 0.3088$\pm$0.0042\uarblank & 0.7051$\pm$0.0108\uarblank & baseline \\
\makecell[l]{WavLM-Large + clap}   & 0.8086$\pm$0.0497\uardown & \bestblock{\textbf{0.3118$\pm$0.0042}}\uarup & \bestblock{0.7142$\pm$0.0201}\uarup & 2$\uparrow$, 1$\downarrow$ \\
\makecell[l]{WavLM-Large + reclap} & 0.8125$\pm$0.0442\uardown & 0.3046$\pm$0.0019\uardown & 0.6939$\pm$0.0209\uardown & 0$\uparrow$, 3$\downarrow$ \\
\makecell[l]{WavLM-Large + clsp}   & \bestblock{\textbf{0.8711$\pm$0.0387}}\uarup & 0.3041$\pm$0.0017\uardown & 0.7111$\pm$0.0074\uarup & 2$\uparrow$, 1$\downarrow$ \\
\midrule
\rowcolor{black!4}\multicolumn{5}{@{}l@{}}{\textbf{FM 2.1: HuBERT-Base}}\\
\makecell[l]{HuBERT-Base + none}   & \bestblock{0.8125$\pm$0.0221}\uarblank & 0.2723$\pm$0.0001\uarblank & 0.6845$\pm$0.0256\uarblank & baseline \\
\makecell[l]{HuBERT-Base + clap}   & 0.7578$\pm$0.0442\uardown & 0.2755$\pm$0.0021\uarup & 0.6711$\pm$0.0135\uardown & 1$\uparrow$, 2$\downarrow$ \\
\makecell[l]{HuBERT-Base + reclap} & 0.7422$\pm$0.0442\uardown & 0.2675$\pm$0.0007\uardown & 0.6839$\pm$0.0012\uardown & 0$\uparrow$, 3$\downarrow$ \\
\makecell[l]{HuBERT-Base + clsp}   & 0.7969$\pm$0.0110\uardown & \bestblock{0.2927$\pm$0.0045}\uarup & \bestblock{0.6907$\pm$0.0213}\uarup & 2$\uparrow$, 1$\downarrow$ \\
\midrule
\rowcolor{black!4}\multicolumn{5}{@{}l@{}}{\textbf{FM 2.2: HuBERT-Large}}\\
\makecell[l]{HuBERT-Large + none}   & 0.8125$\pm$0.0442\uarblank & 0.2775$\pm$0.0002\uarblank & 0.6987$\pm$0.0229\uarblank & baseline \\
\makecell[l]{HuBERT-Large + clap}   & 0.7539$\pm$0.0387\uardown & 0.2760$\pm$0.0162\uardown & 0.7177$\pm$0.0019\uarup & 1$\uparrow$, 2$\downarrow$ \\
\makecell[l]{HuBERT-Large + reclap} & 0.7539$\pm$0.1602\uardown & 0.2716$\pm$0.0051\uardown & 0.6988$\pm$0.0055\uarup & 1$\uparrow$, 2$\downarrow$ \\
\makecell[l]{HuBERT-Large + clsp}   & \bestblock{0.8242$\pm$0.0055}\uarup & \bestblock{0.2895$\pm$0.0029}\uarup & \bestblock{\textbf{0.7184$\pm$0.0092}}\uarup & 3$\uparrow$, 0$\downarrow$ \\
\bottomrule
\end{tabularx}
\end{adjustbox}
\vspace{-3mm}
\end{table*}

\section{ZS-Fuse}
\subsection{Overview}
We illustrate our proposal Zero-shot Fusion (ZS-Fuse) in Fig.~\ref{fig:zsfuse}.
The first branch performs standard fine-tuning of the speech foundation model, yielding a $D$-dimensional vector $\textbf{h}$.
The second branch uses a dual-encoder Audio-Language Model (ALM) to obtain zero-shot predictions ($\textbf{s}$) for each emotion class ($E$).
$P$ prompts are used for querying, and their zero-shot scores are averaged, making this a simplistic version of Zero-Shot Prompt Ensembling~\cite{allingham2023simple}.
Finally, we use LayerNorm (LN) and concatenation to do the late fusion as $\mathbf{z}=\big[\mathbf{h}\,;\,\mathrm{LN}(\mathbf{s})\big]\in\mathbb{R}^{D+E}$, which is then fed to the classification output layer.
ZS-Fuse may be seen as semantic grounding without explicit Automatic Speech Recognition.

\begin{figure*}[t]
    \centering
    \includegraphics[width=0.68\linewidth]{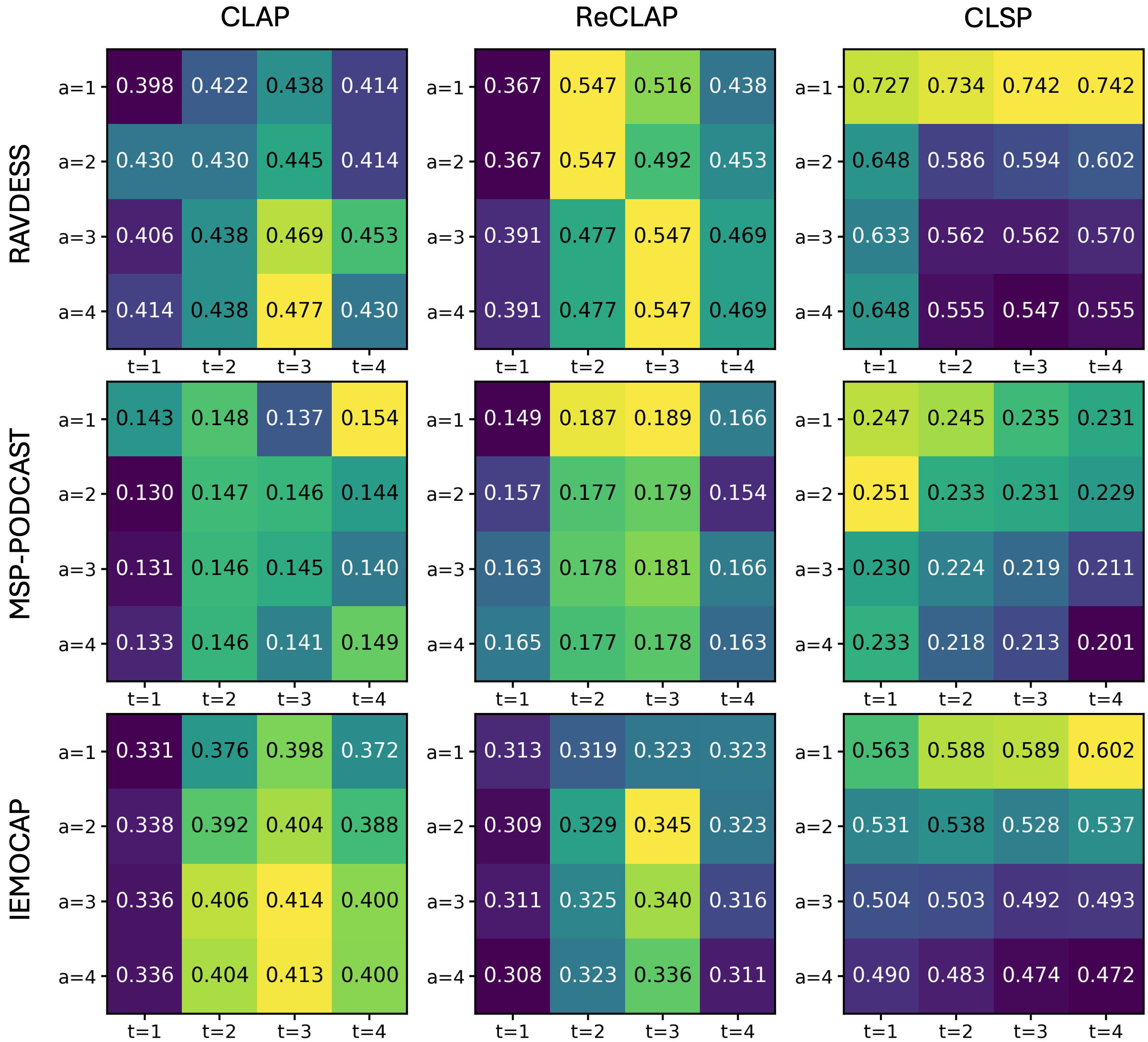}
    \caption{Zero-shot UAR of three ALMs on three datasets for 16 combinations of audio and text repeats ($[1,2,3,4]\times[1,2,3,4]$).}
    \label{fig:zsscores}
    \vspace{-2mm}
\end{figure*}

\subsection{Prompt Engineering}
We use three prompts per emotion class: an emotion+speech form (“angry speech”), a speaker-centric form (“a person speaking angrily”), and a voice-attribute form (“anger in the voice”).
The ensemble covers the classes used in each dataset: A (anger), C (contempt), D (disgust), F (fear), H (happiness), N (neutral), S (sadness), and U (surprise).
A distinct feature of our prompt ensemble is its simplicity.
Prompt engineering is shown to influence zero-shot performance~\cite{zhou2022learning,ghosh2025reclap,olvera2024sound}.
We also propose a technique, \textit{prompt amplification}, which serves as a controlled amplification knob.
In audio, it simply repeats the signal, while in text, it replaces \texttt{<emotion\_j> speech} with \texttt{N instances of <emotion\_j> speech}.
This enables us to safely discover stronger zero-shot capabilities of Audio-Language Models in a controlled manner.
We defer fine-grained prompting~\cite{ghosh2025reclap} and bigger prompt ensembles to future work.

\begin{table*}[t]
\centering
\caption{
Effect of audio/text repetition on UAR (mean$\pm$std).
We report Best/Default ($a{=}1,t{=}1$)/Worst combinations (among total 16), the range $\Delta=\text{Best}-\text{Worst}$,
and the no-ALM baseline.
Per case (A,B), we report best and second best in bold and underline.
}
\label{tab:final}
\setlength{\tabcolsep}{3.2pt}
\renewcommand{\arraystretch}{1.08}
\begin{adjustbox}{max width=0.97\textwidth}
\small
\begin{tabularx}{\textwidth}{@{}l l >{\centering\arraybackslash}X >{\centering\arraybackslash}X >{\centering\arraybackslash}X >{\centering\arraybackslash}X >{\centering\arraybackslash}X@{}}
\toprule
\textbf{System} & \textbf{Dataset} &
\textbf{Best $a\times t$} &
\textbf{Default $1\times1$} &
\textbf{Worst $a\times t$} &
\textbf{Range $\Delta$ (Best-Worst)} &
\textbf{No ALM Baseline} \\
\midrule
\rowcolor{black!4}\multicolumn{7}{@{}l@{}}{\textbf{Case A: WavLM-Large + CLSP}}\\
WavLM-Large + CLSP & RAVDESS &
$a1{\times}t2$: \textbf{0.8984$\pm$0.0221} &
$a1{\times}t1$: \underline{0.8750$\pm$0.0442} &
$a3{\times}t2$: 0.8008$\pm$0.0276 &
0.0977 &
$a0{\times}t0$: 0.8438$\pm$0.0552 \\
WavLM-Large + CLSP & MSP-Podcast &
$a1{\times}t2$: \textbf{0.3108$\pm$0.0008} &
$a1{\times}t1$: 0.3081$\pm$0.0035 &
$a2{\times}t3$: 0.3020$\pm$0.0011 &
0.0088 &
$a0{\times}t0$: \underline{0.3088$\pm$0.0042} \\
WavLM-Large + CLSP & IEMOCAP &
$a1{\times}t4$: \textbf{0.7254$\pm$0.0021} &
$a1{\times}t1$: \underline{0.7118$\pm$0.0031} &
$a3{\times}t3$: 0.7106$\pm$0.0117 &
0.0148 &
$a0{\times}t0$: 0.7051$\pm$0.0108 \\
\midrule
\rowcolor{black!4}\multicolumn{7}{@{}l@{}}{\textbf{Case B: WavLM-Base+ + CLAP}}\\
WavLM-Base+ + CLAP & RAVDESS &
$a4{\times}t2$: \textbf{0.7703$\pm$0.0356} &
$a1{\times}t1$: \underline{0.7422$\pm$0.1038} &
$a2{\times}t2$: 0.6969$\pm$0.0666 &
0.0734 &
$a0{\times}t0$: 0.6953$\pm$0.0994 \\
WavLM-Base+ + CLAP & MSP-Podcast &
$a2{\times}t4$: \underline{0.2947$\pm$0.0009} &
$a1{\times}t1$: 0.2908$\pm$0.0006 &
$a4{\times}t3$: 0.2807$\pm$0.0032 &
0.0140 &
$a0{\times}t0$: \textbf{0.2955$\pm$0.0037} \\
WavLM-Base+ + CLAP & IEMOCAP &
$a1{\times}t4$: \textbf{0.7017$\pm$0.0056} &
$a1{\times}t1$: \underline{0.6833$\pm$0.0078} &
$a1{\times}t3$: 0.6799$\pm$0.0003 &
0.0218 &
$a0{\times}t0$: 0.6736$\pm$0.0060 \\
\bottomrule
\end{tabularx}
\end{adjustbox}
\vspace{-4mm}
\end{table*}

\section{Methodology}
Here, we describe the components of the ZS-Fuse.

\subsection{Test Datasets}
\underline{\textbf{RAVDESS}}~\cite{livingstone2018ryerson}:
This is a validated corpus of emotional speech and song from which we extract speech-only recordings for the 4-class subset: angry, happy, sad, and neutral, with distribution: \{N:96, H:192, S:192, A:192\}.
We create a speaker-disjoint split across the 24 actors, yielding 672 utterances in total, with 448/112/112 train/validation/test samples.
The total dataset is 0.69~h, and the average utterance is 3.70~s.

\underline{\textbf{MSP-PODCAST}}~\cite{busso2025msp}:
This is a large corpus of naturalistic emotional speech drawn from podcast recordings and annotated with both categorical emotion labels and dimensional ratings such as arousal, valence, and dominance.
Class distribution is \{N:52753, H:29454, S:8347, A:10342, U:5227, C:5127, D:2827, F:1997\}.
We use the provided partitions, and evaluate on merged ``test1'' and ``test2''.
The resulting labeled subset contains 116,074 utterances (65,205 train / 15,341 dev / 35,528 test), with a total duration of 182.9~h and an average utterance length of 5.67~s.

\underline{\textbf{IEMOCAP}}~\cite{busso2008iemocap}:
We use IEMOCAP in the standard 4-class setting comprising neutral, sad, angry, and happy, with excited merged with happy.
Class distribution is \{N:1708, H:1636, S:1084, A:1103\}.
We have 5,531 unique utterances from 10 speakers, totaling 6.99~h and an average utterance length of 4.55 s. 
We then construct 5 leave-one-session-out folds, using 3/1/1 sessions for train/val/test, respectively.

\subsection{Foundation Models}
We evaluate WavLM-Base+ (94.7M), WavLM-Large (316.6M)~\cite{chen2022wavlm}, HuBERT-Base (90M), and HuBERT-Large (300M)~\cite{hsu2021hubert}, all of which are pretrained self-supervised speech encoders.
WavLM augments the HuBERT framework with gated relative position bias and denoising-based pretraining, whereas HuBERT uses masked prediction of hidden units.
We investigated Whisper~\cite{radford2023robust} of similar sizes but it gave inferior results.

\subsection{Dual-Encoder Audio-Language Models}
We investigate three ALMs: 1) CLAP (Contrastive Language-Audio Pretraining)~\cite{elizalde2023clap} (specifically the stronger LAION CLAP~\cite{wu2023large} version), 2) ReCLAP~\cite{ghosh2025reclap}, and 3) CLSP (Contrastive Language-Speech Pretraining)~\cite{yang2026towards}.
CLAP is trained with contrastive learning to align audio and text (caption) in a joint embedding space for open-vocabulary retrieval.
ReCLAP improves training by enhancing training captions with a Large Language Model (LLM).
CLSP goes further ahead by incorporating fine-grained supervision via external captioning models and agentic verification for filtering.

\subsection{Training details}
We keep ALM frozen (both its text and speech encoders) and audio FM unfrozen.
Training is performed using the AdamW optimizer, a batch size of 32, 30 epochs, a learning rate of 2e-5, and with 3 random seeds to capture training stochasticity via standard deviation.
The best validation UAR epoch is used to report the test UAR.

\section{Results}

\subsection{Baseline}
In Table~\ref{tab:baseline}, we report Unweighted Average Recall (UAR) for three datasets and four foundation models.
In the first row, we note the random UAR value for reference.
The first row of each result block, ``\texttt{FM + none}'', is the baseline that uses only FM fine-tuning.
Here, we see that WavLM-Large substantially outperforms the other three FMs on all three datasets.
WavLM-Large achieves close to 0.84 UAR on RAVDESS, 0.30 UAR on MSP-PODCAST, and 0.70 UAR on IEMOCAP.
The next three rows utilize the three ALMs.
The last column reports the number of datasets on which improvement ($\uparrow$) or degradation ($\downarrow$) is observed when using the dual encoders.
Across the 12 runs, CLAP improved UAR in 5/12, ReCLAP in 3/12, and CLSP in 9/12. Thus, CLSP showed the most consistent gains relative to the corresponding \texttt{FM + none} baseline.
Note that these results are without prompt amplification ($a1 \times t1$).

\subsection{Comprehensive zero-shot evaluation}
In Fig.~\ref{fig:zsscores}, we note the zero-shot performance of three ALMs with different audio and text repeats.
We first perform a sanity check to ensure that all settings outperform the random baseline.
Visually, we observe that lighter colors (better performance) are toward the right (more text repeats) and downward (more audio repeats).
More repetitions of either modality generally help; however, CLSP degrades with audio repeats, which we attribute to software implementation since we limited the audio length to 6~s due to memory constraints.
Overall, CLSP performs best on RAVDESS (UAR=0.742, $a1\times t4$), even surpassing the WavLM-Base-Plus supervised baseline (UAR=0.695 from Table~\ref{tab:baseline}).
In the literature, more complex ALMs, such as Audio Flamingo 3 (AF3)~\cite{goel2025audio} and Qwen2-A-Inst~\cite{chu2024qwen2}, achieve IEMOCAP performance of 63.8\% and 59.2\%, respectively, whereas CLSP zero-shot performance reaches 60.2\% ($a1\times t4$ config), indicating that dual-encoder ALMs are underutilized.
We also note that zero-shot performance exhibits high variability with respect to the number of modality repeats and shows no clear trends along the two axes, making it difficult to determine which $a\times t$ config to deploy.

\subsection{Results with Prompt Amplification}
In Table~\ref{tab:final}, we report a summary of ZS-Fuse results.
Of the 16 repeat combinations, we present the best, the default (a1xt1), and the worst.
Since we found CLSP as the best ALM in the zero-shot study (Fig.~\ref{fig:zsscores}), we conduct this study using WavLM-Large (Case A).
For completeness, we also present WavLM-Base-Plus with CLAP (Case B).
Observations are similar in both cases, with improvements in all datasets except MSP-PODCAST (case B), which is more suited for processing by CLAP instead of CLSP, as suggested by Fig.~\ref{fig:zsscores}.
Except for one result, the best $a \times t$ substantially improves over the default $a1\times t1$ as well as the no ALM baseline.
The worst $a \times t$ counterintuitively degrades performance, sometimes even below the no ALM baseline.
This shows that the prompt amplification trick can have unexpected behavior during deployment.
Thus, our proposal is powerful for achieving state-of-the-art results, but the prompt amplification component remains a discrete optimization strategy.
Prompt amplification can also be seen as a test-time self-conditioning~\cite{kataria2023self} / ensembling~\cite{allingham2023simple} trick with potential bias-correction and calibration effects~\cite{jiang2023calibrating}.
In future work, we will investigate this further and work on establishing a 1:1 correspondence between isolated zero-shot strength and downstream utility to enhance the reliability of ZS-Fuse.

\section{Conclusion}
We introduced ZS-Fuse, a late-fusion method that safely integrates zero-shot speech emotion predictions derived from ALMs into the conventional \textit{supervised foundation models} pipeline.
We establish the efficacy of our proposal by exploring three dual-encoder ALMs (CLAP, ReCLAP, CLSP), two FMs (WavLM, HuBERT), and three datasets (RAVDESS, MSP-PODCAST, IEMOCAP).
To discover stronger zero-shot capabilities, we devise a novel technique, \textit{prompt amplification}, that repeats the audio and text queries.
Finally, we show that default (no audio or text repeat) and best-repeat combinations can substantially surpass the state-of-the-art.
In future work, we will improve prompt amplification and extend the study to generative ALMs and better prompt ensembles.

\clearpage

\bibliographystyle{unsrt}  
\bibliography{main}        

@article{goel2025audio,
  title={Audio flamingo 3: Advancing audio intelligence with fully open large audio language models},
  author={Goel, Arushi and Ghosh, Sreyan and Kim, Jaehyeon and Kumar, Sonal and Kong, Zhifeng and Lee, Sang-gil and Yang, Chao-Han Huck and Duraiswami, Ramani and Manocha, Dinesh and Valle, Rafael and others},
  journal={arXiv preprint arXiv:2507.08128},
  year={2025}
}

@article{chu2024qwen2,
  title={Qwen2-audio technical report},
  author={Chu, Yunfei and Xu, Jin and Yang, Qian and Wei, Haojie and Wei, Xipin and Guo, Zhifang and Leng, Yichong and Lv, Yuanjun and He, Jinzheng and Lin, Junyang and others},
  journal={arXiv preprint arXiv:2407.10759},
  year={2024}
}

@article{zhao2025steering,
  title={Steering language model to stable speech emotion recognition via contextual perception and chain of thought},
  author={Zhao, Zhixian and Zhu, Xinfa and Wang, Xinsheng and Wang, Shuiyuan and Geng, Xuelong and Tian, Wenjie and Xie, Lei},
  journal={IEEE Transactions on Audio, Speech and Language Processing},
  volume={34},
  pages={415--426},
  year={2025},
  publisher={IEEE}
}

@article{chen2022wavlm,
  title={Wavlm: Large-scale self-supervised pre-training for full stack speech processing},
  author={Chen, Sanyuan and Wang, Chengyi and Chen, Zhengyang and Wu, Yu and Liu, Shujie and Chen, Zhuo and Li, Jinyu and Kanda, Naoyuki and Yoshioka, Takuya and Xiao, Xiong and others},
  journal={IEEE Journal of Selected Topics in Signal Processing},
  volume={16},
  number={6},
  pages={1505--1518},
  year={2022},
  publisher={IEEE}
}

@article{zhou2022learning,
  title={Learning to prompt for vision-language models},
  author={Zhou, Kaiyang and Yang, Jingkang and Loy, Chen Change and Liu, Ziwei},
  journal={International journal of computer vision},
  volume={130},
  number={9},
  pages={2337--2348},
  year={2022},
  publisher={Springer}
}

@inproceedings{ghosh2025reclap,
  title={Reclap: Improving zero shot audio classification by describing sounds},
  author={Ghosh, Sreyan and Kumar, Sonal and Evuru, Chandra Kiran Reddy and Nieto, Oriol and Duraiswami, Ramani and Manocha, Dinesh},
  booktitle={ICASSP 2025-2025 IEEE International Conference on Acoustics, Speech and Signal Processing (ICASSP)},
  pages={1--5},
  year={2025},
  organization={IEEE}
}

@article{olvera2024sound,
  title={A sound description: Exploring prompt templates and class descriptions to enhance zero-shot audio classification},
  author={Olvera, Michel and Stamatiadis, Paraskevas and Essid, Slim},
  journal={arXiv preprint arXiv:2409.13676},
  year={2024}
}

@inproceedings{allingham2023simple,
  title={A simple zero-shot prompt weighting technique to improve prompt ensembling in text-image models},
  author={Allingham, James Urquhart and Ren, Jie and Dusenberry, Michael W and Gu, Xiuye and Cui, Yin and Tran, Dustin and Liu, Jeremiah Zhe and Lakshminarayanan, Balaji},
  booktitle={International Conference on Machine Learning},
  pages={547--568},
  year={2023},
  organization={PMLR}
}

@article{kataria2023self,
  title={Self-FiLM: Conditioning GANs with self-supervised representations for bandwidth extension based speaker recognition},
  author={Kataria, Saurabh and Villalba, Jes{\~A}{\textordmasculine}s and Thebaud, Thomas and Dehak, Najim and others},
  journal={arXiv preprint arXiv:2303.03657},
  year={2023}
}

@article{jiang2023calibrating,
  title={Calibrating language models via augmented prompt ensembles},
  author={Jiang, Mingjian and Ruan, Yangjun and Huang, Sicong and Liao, Saifei and Pitis, Silviu and Grosse, Roger Baker and Ba, Jimmy},
  year={2023}
}

@article{livingstone2018ryerson,
  title={The Ryerson Audio-Visual Database of Emotional Speech and Song (RAVDESS): A dynamic, multimodal set of facial and vocal expressions in North American English},
  author={Livingstone, Steven R and Russo, Frank A},
  journal={PloS one},
  volume={13},
  number={5},
  pages={e0196391},
  year={2018},
  publisher={Public Library of Science}
}

@article{busso2025msp,
  title={The msp-podcast corpus},
  author={Busso, Carlos and Lotfian, Reza and Sridhar, Kusha and Salman, Ali N and Lin, Wei-Cheng and Goncalves, Lucas and Parthasarathy, Srinivas and Naini, Abinay Reddy and Leem, Seong-Gyun and Martinez-Lucas, Luz and others},
  journal={arXiv preprint arXiv:2509.09791},
  year={2025}
}

@article{busso2008iemocap,
  title={IEMOCAP: Interactive emotional dyadic motion capture database},
  author={Busso, Carlos and Bulut, Murtaza and Lee, Chi-Chun and Kazemzadeh, Abe and Mower, Emily and Kim, Samuel and Chang, Jeannette N and Lee, Sungbok and Narayanan, Shrikanth S},
  journal={Language resources and evaluation},
  volume={42},
  number={4},
  pages={335--359},
  year={2008},
  publisher={Springer}
}

@article{hsu2021hubert,
  title={Hubert: Self-supervised speech representation learning by masked prediction of hidden units},
  author={Hsu, Wei-Ning and Bolte, Benjamin and Tsai, Yao-Hung Hubert and Lakhotia, Kushal and Salakhutdinov, Ruslan and Mohamed, Abdelrahman},
  journal={IEEE/ACM transactions on audio, speech, and language processing},
  volume={29},
  pages={3451--3460},
  year={2021},
  publisher={IEEE}
}

@inproceedings{elizalde2023clap,
  title={Clap learning audio concepts from natural language supervision},
  author={Elizalde, Benjamin and Deshmukh, Soham and Al Ismail, Mahmoud and Wang, Huaming},
  booktitle={ICASSP 2023-2023 IEEE International Conference on Acoustics, Speech and Signal Processing (ICASSP)},
  pages={1--5},
  year={2023},
  organization={IEEE}
}

@article{yang2026towards,
  title={Towards Fine-Grained and Multi-Granular Contrastive Language-Speech Pre-training},
  author={Yang, Yifan and Han, Bing and Wang, Hui and Wang, Wei and Ma, Ziyang and Zhou, Long and Jin, Zengrui and Yang, Guanrou and Wang, Tianrui and Tan, Xu and others},
  journal={arXiv preprint arXiv:2601.03065},
  year={2026}
}

@inproceedings{wu2023large,
  title={Large-scale contrastive language-audio pretraining with feature fusion and keyword-to-caption augmentation},
  author={Wu, Yusong and Chen, Ke and Zhang, Tianyu and Hui, Yuchen and Berg-Kirkpatrick, Taylor and Dubnov, Shlomo},
  booktitle={ICASSP 2023-2023 IEEE International Conference on Acoustics, Speech and Signal Processing (ICASSP)},
  pages={1--5},
  year={2023},
  organization={IEEE}
}

@inproceedings{schuller2013interspeech,
  title={The INTERSPEECH 2013 computational paralinguistics challenge: Social signals, conflict, emotion, autism},
  author={Schuller, Bj{\"o}rn and Steidl, Stefan and Batliner, Anton and Vinciarelli, Alessandro and Scherer, Klaus and Ringeval, Fabien and Chetouani, Mohamed and Weninger, Felix and Eyben, Florian and Marchi, Erik and others},
  booktitle={Proceedings INTERSPEECH 2013, 14th Annual Conference of the International Speech Communication Association, Lyon, France},
  year={2013}
}

@article{ma2020survey,
  title={A survey on empathetic dialogue systems},
  author={Ma, Yukun and Nguyen, Khanh Linh and Xing, Frank Z and Cambria, Erik},
  journal={Information Fusion},
  volume={64},
  pages={50--70},
  year={2020},
  publisher={Elsevier}
}

@article{parra2022classification,
  title={Classification of emotions and evaluation of customer satisfaction from speech in real world acoustic environments},
  author={Parra-Gallego, Luis Felipe and Orozco-Arroyave, Juan Rafael},
  journal={Digital Signal Processing},
  volume={120},
  pages={103286},
  year={2022},
  publisher={Elsevier}
}

@article{jordan2025speech,
  title={Speech emotion recognition in mental health: Systematic review of voice-based applications},
  author={Jordan, Eric and Terrisse, Rapha{\"e}l and Lucarini, Valeria and Alrahabi, Motasem and Krebs, Marie-Odile and Descl{\'e}s, Julien and Lemey, Christophe},
  journal={JMIR mental health},
  volume={12},
  number={1},
  pages={e74260},
  year={2025},
  publisher={JMIR Publications Inc., Toronto, Canada}
}

@inproceedings{ma2023emotion,
  title={Emotion-aware voice assistants: Design, implementation, and preliminary insights},
  author={Ma, Yong and Zhang, Yuchong and Bachinski, Miroslav and Fjeld, Morten},
  booktitle={Proceedings of the Eleventh International Symposium of Chinese CHI},
  pages={527--532},
  year={2023}
}

@article{ma2024emobox,
  title={Emobox: Multilingual multi-corpus speech emotion recognition toolkit and benchmark},
  author={Ma, Ziyang and Chen, Mingjie and Zhang, Hezhao and Zheng, Zhisheng and Chen, Wenxi and Li, Xiquan and Ye, Jiaxin and Chen, Xie and Hain, Thomas},
  journal={arXiv preprint arXiv:2406.07162},
  year={2024}
}

@article{zhang2026voxemo,
  title={VoxEmo: Benchmarking Speech Emotion Recognition with Speech LLMs},
  author={Zhang, Hezhao and Chou, Huang-Cheng and Narayanan, Shrikanth and Hain, Thomas},
  journal={arXiv preprint arXiv:2603.08936},
  year={2026}
}

@article{huang2025survey,
  title={A survey on hallucination in large language models: Principles, taxonomy, challenges, and open questions},
  author={Huang, Lei and Yu, Weijiang and Ma, Weitao and Zhong, Weihong and Feng, Zhangyin and Wang, Haotian and Chen, Qianglong and Peng, Weihua and Feng, Xiaocheng and Qin, Bing and others},
  journal={ACM Transactions on Information Systems},
  volume={43},
  number={2},
  pages={1--55},
  year={2025},
  publisher={ACM New York, NY}
}

@article{xu2025qwen3,
  title={Qwen3-omni technical report},
  author={Xu, Jin and Guo, Zhifang and Hu, Hangrui and Chu, Yunfei and Wang, Xiong and He, Jinzheng and Wang, Yuxuan and Shi, Xian and He, Ting and Zhu, Xinfa and others},
  journal={arXiv preprint arXiv:2509.17765},
  year={2025}
}

@article{dutta2025abhinaya,
  title={ABHINAYA--A System for Speech Emotion Recognition In Naturalistic Conditions Challenge},
  author={Dutta, Soumya and Balaji, Smruthi and Salinamakki, Viveka and Ganapathy, Sriram and others},
  journal={arXiv preprint arXiv:2505.18217},
  year={2025}
}

@inproceedings{ma2024emotion2vec,
  title={emotion2vec: Self-supervised pre-training for speech emotion representation},
  author={Ma, Ziyang and Zheng, Zhisheng and Ye, Jiaxin and Li, Jinchao and Gao, Zhifu and Zhang, Shiliang and Chen, Xie},
  booktitle={Findings of the Association for Computational Linguistics: ACL 2024},
  pages={15747--15760},
  year={2024}
}

@article{jing2024paraclap,
  title={ParaCLAP--Towards a general language-audio model for computational paralinguistic tasks},
  author={Jing, Xin and Triantafyllopoulos, Andreas and Schuller, Bj{\"o}rn},
  journal={arXiv preprint arXiv:2406.07203},
  year={2024}
}

@inproceedings{radford2023robust,
  title={Robust speech recognition via large-scale weak supervision},
  author={Radford, Alec and Kim, Jong Wook and Xu, Tao and Brockman, Greg and McLeavey, Christine and Sutskever, Ilya},
  booktitle={International conference on machine learning},
  pages={28492--28518},
  year={2023},
  organization={PMLR}
}

\end{document}